\documentstyle[aps,pre,psfig,floats]{revtex}

\begin{document}                             
\newcommand{\C}{{\cal C}}
\newcommand{\G}{{\cal G}}

\psfigurepath{/users3/guyh/spin_glasses/letter}

\twocolumn[\hsize\textwidth\columnwidth\hsize\csname@twocolumnfalse%
\endcsname

\title{Spin Domains Generate Hierarchical Ground State 
Structure in $J=\pm 1$ Spin Glasses}
\author{Guy Hed$^1$, Alexander K. Hartmann$^2$,
Dietrich Stauffer$^3$ and Eytan Domany$^1$\\[2mm]}
\address{$^{1}$ Department of Physics of Complex Systems,
    Weizmann Institute of Science, Rehovot 76100, Israel\\
    $^{2}$ Institut f\"ur Theoretische Physik, Universit\"at G\"ottingen,
    Bunsenstr. 9, 37073 G\"ottingen, Germany \\
    $^{3}$ Institute for Theoretical Physics, University of Cologne,  
    Z\"ulpicher Stra{\ss}e 77, D-50937 K\"oln, Germany }
\date{\today}
\maketitle

\begin{abstract}
Unbiased samples of ground states were generated for the short-range
Ising spin glass with $J_{ij}=\pm 1$, in three dimensions.
Clustering the ground states revealed their hierarchical structure, which
is explained by correlated spin domains,
serving as cores for macroscopic zero energy ``excitations''.
\end{abstract}

]

Mean-field (MF) theory~\cite{Mezard84} provides a commonly
accepted  description of the low $T$ phase of infinite-range spin glasses
~\cite{SK75}. It
predicts many pure states with an hierarchical ultrametric
organization 
and a non-trivial state overlap distribution
$P(q)$. Although this structure
was suggested to hold also for short-range spin glasses 
(SRSG)~\cite{Marinari98},  
the equilibrium properties of these  are still
controversial. The main dispute concerns the number
of different thermodynamic (pure) states of the system
below $T_c$.  
Fisher and Huse~\cite{Fisher88} studied SRSG 
with continuously distributed couplings. 
According to them, a finite region embedded in an infinite system
will be in one of two pure states. They 
describe the system's low energy
excitations above the local ground state as {\em finite} flipped 
spin domains. 
By flipping compact {\em macroscopic} domains one can generate other
pure states. 
This structure of pure states (which we call the FH scenario) yields,
{\it for any finite region of the infinite system},
a trivial $P(q)$ distribution. 
Numerical evidence for non-trivial $P(q)$ in {\it finite} systems
~\cite{Krzakala00,Katzgraber00,Franz00b} 
does {\it not} contradict the FH scenario~\cite{Huse87}.

We present here evidence for a new picture of the spin glass phase.
{\it It
possesses some characteristics of the MF description, such as
non-trivial $P(q)$ and a hierarchical} (but non-ultrametric!) 
{\it structure of the pure
states; nevertheless, it is also consistent with the FH scenario.}

{\it The model:}
We study the ground states (GS) of
the Edwards-Anderson model~\cite{EA75} of an Ising spin glass
\begin{equation}
{\cal H} = \sum_{\langle ij \rangle}J_{ij}S_iS_j ~, \qquad 	J_{ij}=\pm 1
\label{eq:H}
\end{equation}
$\langle ij \rangle$ denotes nearest neighbor sites of a simple cubic lattice;
the  values $J_{ij}=\pm 1$ are 
assigned to each bond independently and with
equal probabilities\cite{foot1}.
Although this model is very special - it has highly  
degenerate GS~\cite{Kirkpatrick77} - we expect that
its low-$T$ properties are  generic, i.e.
not qualitatively different from other $\{ J_{ij} \}$ distributions
(such as Gaussian). On the other hand, the low-$T$ properties of 
(\ref{eq:H})
are most probably dominated by the structure of its GS.
Hence we hope that 
the GS of the $J=\pm 1$ model
provide insight about the low-$T$ behavior of generic 
short-range Ising spin glass. In any case,
the GS structure of this model is  interesting
on its own merit.

By combining very efficient 
algorithms~\cite{Hartmann96} that produce ground states of 
this model, with simulated tempering (ST)~\cite{Kerler94}, we 
generated {\it unbiased} samples 
of the GS; i.e. we ``equilibrated" our
system at $T=0$. We studied  the model (\ref{eq:H}) 
with periodic boundary conditions
in 3 dimensions, with $N=L^3$ spins, for $L=4,5,6,8$. 
For each size $L$ we produced
800 to 1000 realizations $\{ J \}$; 
for each realization an unbiased
sample of $M = 500$ GS has been generated and analyzed.

\begin{figure}[t]
\centerline{\psfig{figure=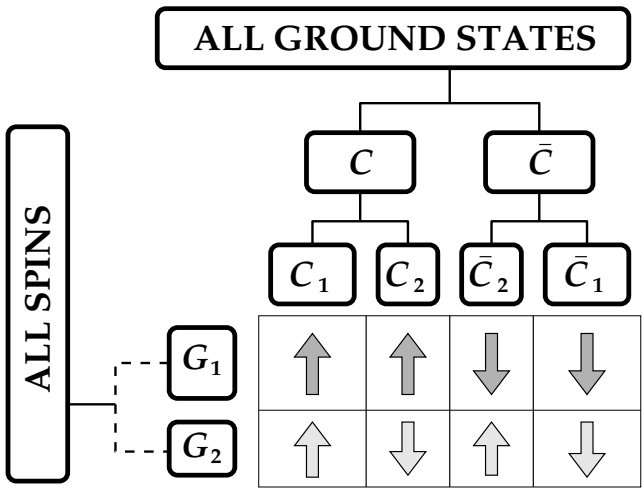,width=85mm}}
\vspace{2mm}
\caption{Schematic representation of our picture; the two largest
spin domains and the first two levels in 
the hierarchical organization of the GS are shown. 
The structure of the GS is explained by the
spin domains' orientations; e.g.
in the GS of the two sets $\C_1,\C_2$, the spins of $\G_1$ have the same
orientation,  whereas the spins of the smaller 
cluster, $\G_2$, have flipped. }
\label{Fig1}
\end{figure}

\begin{figure}[t]
\centerline{\psfig{figure=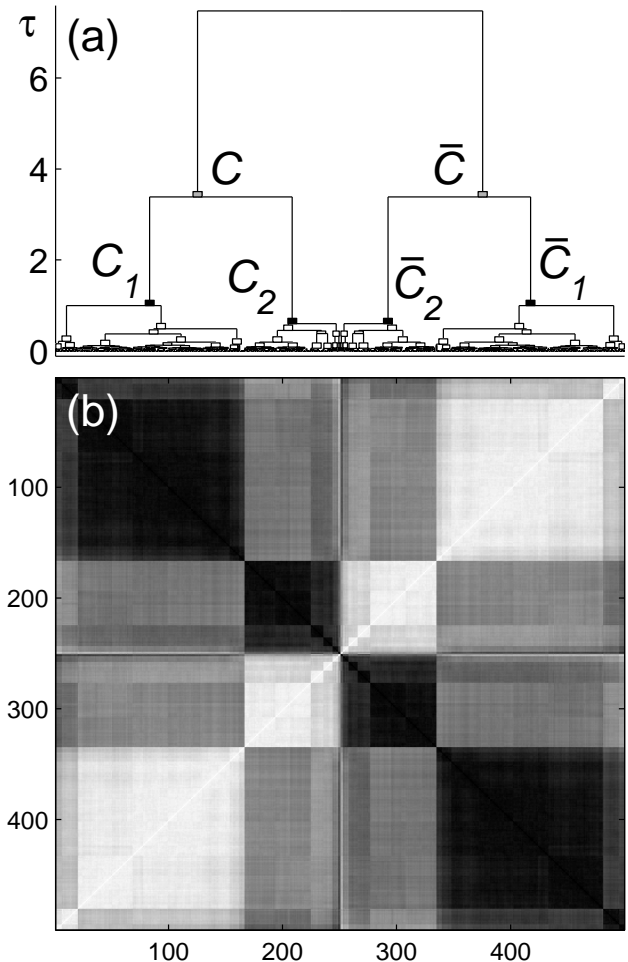,width=8.5cm}}
\caption{{\bf (a)} 
The dendrogram obtained by clustering the ground states, for a particular
set of $J_{ij}$, for $N=6^3$ spins. {\bf (b)} When 
the states are ordered according to the dendrogram,
a clear block structure is seen in $D_{\mu \nu}$,
the distance matrix of the GS. 
Darker shades correspond to shorter distances.}
\label{fig:state_dend}
\end{figure}

{\it Summary of the main results:}
1. For any given $\{ J \}$, the GS  
do {\it not} cover the hypercube ${\mathbf S}=(S_1,S_2,...S_N)$ 
uniformly; rather, there is a  {\it hierarchical structure},
as shown schematically in Fig. \ref{Fig1} and in detail in
\ref{fig:state_dend}.
The set of all GS splits into two {\bf state clusters} $\C$ and 
$\bar \C$, related by {\it spin reversal}; $\C$ splits into  $\C_1$
and $\C_2$, and so on.
 
2. This structure is generated by domains
of highly correlated spins~\cite{Barahona82},
with very different sizes. Separation of GS into $\C$ and $\bar \C$
is determined by the largest {\bf spin domain} $\G_1$,
whose size is typically larger
than $N/2$. Two states in which $\G_1$ has the same 
orientation have a much
higher overlap than two states in which the spins of $\G_1$ are
inverted. This implies formation of two clusters of states, $\C$ and
$\bar \C$, corresponding to the two possible orientations of $\G_1$.
The same structure persists at the next level, 
where the second largest {\bf spin domain}
$\G_2$ determines the partition of $\C$ into $\C_1$ and $\C_2$ (see Fig.
\ref{Fig1}). 
Note that $\G_1$ has the
same orientation in these two clusters.

3. The hierarchical structure of the GS is due to 
large differences between the sizes of the spin domains (typically
$|\G_1|>4|\G_2|$). For domains of equal sizes
no hierarchy would have been observed.

4. This picture {\it differs} from MF; the correlated
domains determine the overlap distribution, and the GS  do not exhibit
ultrametricity. 
On the other hand, {\em if} in the
$L \rightarrow \infty$ limit all but a vanishing
fraction of the spins belong to compact~\cite{Palassini00}
macroscopic correlated domains $\G_a$, then
any finite region of the infinite       
system will exhibit a trivial $P(q)$, in agreement with the FH scenario
(we have not tested the compactness of $\G_a$). 

We explain  how this picture of GS
structure and spin clusters has been found; we present evidence that 
substantiates our findings, investigate their dependence on system size  
and discuss their implications.

{\it Generating unbiased samples of ground states:}
For every realization $\{ J_{ij} \}$ 
we used the {\em genetic cluster exact approximation} (CEA) algorithm
\cite{Hartmann96} to sample the GS. 
Samples obtained by CEA are, however, biased~\cite{Sandvik99}, 
over weighting GS from
small {\it valleys} (a valley $V$ consists of all the GS that can be 
traversed flipping one spin at a time)
- therefore the probability to miss a valley is {\it lower} than that of
an unbiased method. 
We used three methods to overcome this
bias. For $L=4,5$ and 6 we succeeded, for most realizations $\{ J \}$, 
to enumerate exactly all GS within each
valley; selecting at random  $M$ of the  GS
ensures that
each valley is represented according to its size. 
For some $\{ J \}$
enumeration was not possible; then we used ST~\cite{Kerler94} to generate 
samples
(our Glauber dynamics had a decorrelation time of 
less than two sampling periods for each spin).
For $L=8$ we estimated the size of each valley by
the  method of~\cite{Hartmann00} 
and generated a sample of GS in a valley by a Metropolis MC procedure.
To test that this method~\cite{Hartmann00} indeed yields unbiased GS, 
we sampled 60 realizations 
with ST, and ascertained that for the quantities of significance for
our claims (the size of $\G_2$ and the average correlation $\bar c_{12}$ of
the domains - see below) the estimates  
obtained by the two ways did not differ significantly and systematically. 

{\it Clustering methodology:} Clustering is a powerful way to perform
exploratory analysis of all kinds of data. In general, one 
calculates a {\it distance 
matrix} $d_{ij}$ between the $i=1,...n$ data points,
and determines the underlying hierarchy
of partitions (clusters) in the data.
We used Ward's	
agglomerative algorithm \cite{Jain88}; it 
starts with each data point as a separate cluster and at each step
fuses 
the pair of clusters $\alpha,\beta$ that are
at the shortest effective distance
$\rho_{\alpha \beta}$
from each other, stopping when
all points are in 
one cluster.
Initially $\rho_{\alpha \beta}=d_{\alpha \beta}$ 
; when two clusters $\alpha,\beta$ are fused, the distances 
of the new cluster $\alpha '$ to all unchanged clusters 
$\gamma \neq \alpha,\beta$ are updated~\cite{Jain88}:
\begin{equation}
\rho_{\alpha ' \gamma}=\frac{(n_\alpha+ n_\gamma) \rho_{\alpha \gamma}+
(n_\beta + n_\gamma)
\rho_{\beta \gamma} -n_\gamma \rho_{\alpha \beta}}{n_\alpha+n_\beta+n_\gamma}
\end{equation}
where 
$n_x$ is the number of points in cluster $x$.
The algorithm produces a dendrogram such as Fig.
\ref{fig:state_dend}(a). 	
Leaves  represent individual data points. 
The boxes at the nodes represent 
clusters; they are ordered 
horizontally  in a way that reflects their proximity.
The vertical coordinate of cluster $\alpha^\prime$ is
$\tau (\alpha^\prime)
=\rho_{\alpha\beta}$, i.e. 
{\bf the effective distance} between the two clusters that
were fused to form $\alpha^\prime$.
For the initial (single state) clusters we set $\tau=0$. 
When we fuse two ``natural" clusters, whose separation exceeds significantly
their linear extent, 
the branch {\it above} them is long.  Hence we can identify ``natural"
sub-partitions of a cluster;	
as evident from Fig. \ref{fig:state_dend}(a), $C_1, C_2$ are such
natural sub-partitions of $C$, and also $C, {\bar C}$ are natural
sub-clusters of the set of all GS. 
For each realization we analyzed the data in {\it two ways}: 
1. viewing the GS as $M$
data-points
and 2. viewing the spins as $N$ data-points.

{\it Clustering the Ground States:}
Define $D_{\mu \nu} = (1-q_{\mu \nu})/2$ the {\bf distance between states}
${\mathbf S}^\mu$ and ${\mathbf S}^\nu$;
$q_{\mu \nu}=\sum_i S_i^\mu S_i^\nu/N$ is their overlap.
The dendrogram obtained by clustering $500$ GS
for a  system with
$6^3$ spins is shown in
Fig \ref{fig:state_dend}(a). Hierarchical GS structure 
is evident. To provide a quantitative measure of the extent
to which a partition (of, say, $\C$ to $\C_1$ and $\C_2$)
is ``natural", we evaluate the average distance {\it between}
points in $\C$ and $\bar\C$,
\begin{equation}
D(\C,\bar\C)= \frac{1}{|\C| |\bar\C| }
\sum_{\mu \in \C} \sum_{\nu \in \bar\C} D_{\mu \nu}
\label{eq:DCC}
\end{equation}
and compare it to $D(\C)$, the average distance {\it within} $\C$.
In the same manner we define $D(\C_1,\C_2)$, $D(\C_1)$ and $D(\C_2)$.
For $L=6$ we obtained $D(\C)=0.094$(mean)$\pm 0.067$(s.d.);
$D(\C,\bar\C)=0.906\pm0.067$; $D(\C_1)=0.057\pm0.045$;
$D(\C_2)=0.058\pm0.036$; $D(\C_1,\C_2)=0.178\pm0.143$.
For $L=8$ we obtained $D(\C)=0.078\pm 0.050$;
$D(\C,\bar\C)=0.921\pm0.050$; $D(\C_1)=0.049\pm0.028$;
$D(\C_2)=0.049\pm0.020$; $D(\C_1,\C_2)=0.162\pm0.135$.
The results clearly 
show that the hierarchical structure is real, and
not a mere artifact of Ward's algorithm.

A striking  demonstration of this point can be seen
in Fig. \ref{fig:state_dend}(b), which shows the 
distance matrix $D_{\mu \nu}$ between the GS that were ordered in  
Fig. \ref{fig:state_dend}(a).
Dark represents short distances and light - high.
If we cluster states with $S_i=\pm 1$ at random, 
the reordered distance matrix
is a  greyish square.
Only when the clustered states form a real, well defined hierarchy, does the
reordered distance matrix reveal 
the  structure 
so clearly seen in 
Fig. \ref{fig:state_dend}(b).
To understand this hierarchy of GS we 
investigated the organization of the  $N$ {\it spins} in the
$M$ GS.

Fig. \ref{fig:state_dend}(b) resembles the state distance matrix of the MF
picture. There is, however, one crucial difference. In the MF scenario the
off-diagonal sub-matrices of the distance matrix are {\em uniform}, which
leads to ultrametricity \cite{Mezard84}. For example,
if the {\bf sub-matrix} $\tilde D_{ij}$ for $i\in\C_1$ and $j\in\C_2$ is
uniform, the {\bf width} $w(\C_1,\C_2)$ of the distribution
$P(\tilde D_{ij})$ 
should vanish as $L\rightarrow\infty$. We performed a fit of the form
$w(\C_1,\C_2,L) = w_\infty+A L^{-y}$,
with $A$ and $y$ as fit parameters. The minimum of $\chi^2=6.7\times10^{-7}$
was found for $w_\infty=0.025(2)$ with $y=3.4(8)$. Setting
$w_\infty=0$ we get a worse fit, with $\chi^2=1.9\times10^{-5}$.
Our data supports a non-vanishing value of $w_\infty$, in disagreement
with the ultrametricity of the MF picture.

\begin{figure}[t]
\centerline{\psfig{figure=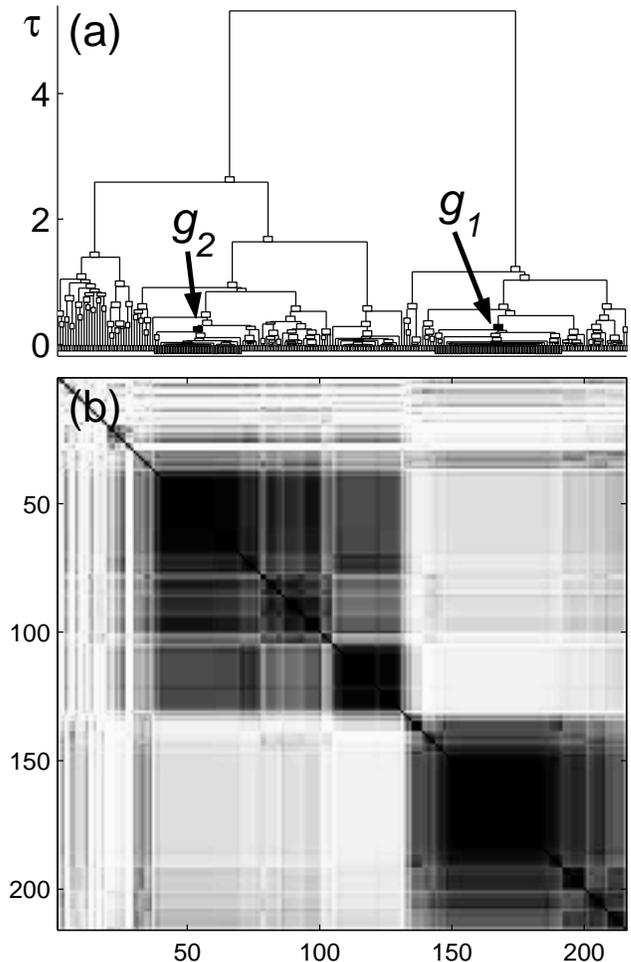,width=85mm}}
\caption{{\bf (a)} The dendrogram $\cal D$ obtained by clustering
the spins of the system of Fig. \ref{fig:state_dend}. For this realization
$g_a=\G_a$ for both $a=1,2$.
{\bf (b)} When the spins are ordered according to the dendrogram,
a structure of correlated spin domains emerges;
darker shades correspond to shorter distances and higher correlations.}
\label{fig:spin_dend}
\end{figure}

{\it Clustering the spins:} 
We cluster $i=1,...,N$
spin-vectors ${\mathbf S}_i=(S_i^1,S_i^2,...,S_i^M)$, 
looking for correlated 
spin domains. 
Define a {\bf distance between spins} $i$ and $j$ by  
$d_{i j} = 1 - {c_{i j}}^2$ ,
where $c_{i j}={\mathbf S}_i \cdot {\mathbf S}_j/M$
is the correlation between spins $i$ and $j$.
Note that ${c_{ij}}^2$ is the relevant measure of correlations in a spin glass.
A typical outcome of clustering the spins with this
distance matrix is the dendrogram $\cal D$ of 
Fig.  \ref{fig:spin_dend}(a),
obtained for the same system, whose GS were studied in 
Fig. \ref{fig:state_dend}.
In Fig. \ref{fig:spin_dend}(b) we show the distance matrix 
obtained {\it after} the spins have been reordered according to 
$\cal D$.
Non-trivial structure in spin space is evident;
dark squares along the diagonal represent highly correlated clusters.

{\it Identifying $\G_1$ and $\G_2$.} 
In order to identify $\G_1$, the largest domain of correlated 
spins,  we go over all pairs of GS, $\mu,\nu$, with  $\mu \in \C$ and 
$\nu \in {\bar \C}$, and identify $\G_{\mu \nu}$, the set of spins that 
have opposite signs in $\mu$ and $\nu$. 
For all $L$ and $\{ J \}$, the set $\G_{\mu \nu}$ is contiguous
for more than 99.5\% of the pairs. Thus, $\G_{\mu \nu}$ can be related to
the low energy excitations found for SRSG with Gaussian
couplings \cite{Krzakala00,Palassini00}.

\begin{figure}[t]
\centerline{
\psfig{figure=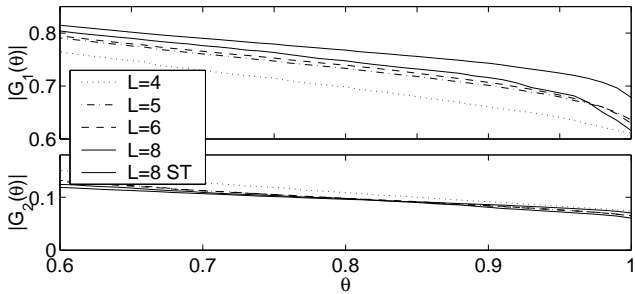,width=85mm}}
\caption{The average sizes of $\G_1(\theta)$ and $\G_2(\theta)$. 
The thin solid line presents only
ST data for $L=8$, which
are unbiased but highly noisy due to the relatively small number of (60) 
realizations.}
\label{fig:gth}
\end{figure}

For a given $\{ J \}$ we identify as
$\G_1$ 
the largest contiguous group of spins shared by at least a fraction 
$\theta = 0.95$ 
of the sets $\G_{\mu\nu}$. Thus, inside $\G_1$ ${c_{ij}}^2\geq0.81$
(for $L=8$ the average correlation inside $\G_1$ is always larger 0.94).
The second largest spin domain, $\G_2$,
is found in a similar way, by scanning all pairs of GS with
$\mu \in \C_1$ and $\nu \in \C_2$. As seen in 
Fig. \ref{fig:gth}, the average sizes of $\G_1, \G_2$ do not change a lot
for $0.60<\theta<0.95$. 

\begin{figure}[t]
\centerline{
\psfig{figure=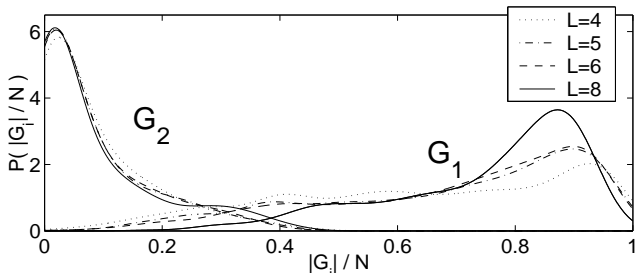,width=85mm}}
\caption{Size distribution of $\G_1$ and $\G_2$. Note that the distribution
of $\G_2$ seems to converge already for $L=8$. The distribution of $\G_1$
converges for sizes between $0.5N$ and $0.7N$.}
\label{fig:g1g2}
\end{figure}

According to our picture we expect $\vert \G_a \vert \propto L^d$
for both $a=1,2$. We present the distributions of $\vert \G_a \vert /L^3$ for
$4 \leq L \leq 8$ in Fig. \ref{fig:g1g2}. Our results show
that the sizes of the two largest spin clusters do scale as $L^3$.

We turn now to identify those clusters $g_1, g_2$ in our spin dendrogram,
which can be  associated  most naturally with the domains  
$\G_1, \G_2$.   
$g_1$ is that 
cluster which is most similar to $\G_1$, i.e. has the largest 
fraction of shared spins ${\cal S}(g_1,\G_1)=2|g_1\cap \G_1|/(|g_1|+|\G_1|)$.
The similarity is high: for $L=8$ on the average we have
${\cal S}(g_a,\G_a)=0.99(1)$ for $a=1$ and 0.97(4) for $a=2$.
This means that $\G_1$ and $\G_2$ do have a meaningful role in spin space,
and are not just an artifact of our analysis.

{\it Overlap distribution:}  

Fig. \ref{fig:g1g2} strongly suggest that $\G_2$ does {\em not} vanish
as $L$ increases, as one can conclude from \cite{Krzakala00}.
Still, the distribution may become trivial if 
\begin{equation}
\bar c_{12} = {1\over|\G_1||\G_2|} 
\sum_{i\in\G_1} \sum_{j\in\G_2} {c_{ij}}^2 \rightarrow 1 ~~~{\rm as}~~L
\rightarrow \infty
\end{equation}
In this case $\G_1$ and $\G_2$ will
always flip together. We carried out fits of the form
$\bar c_{12}(L) = \bar c_{12}(\infty) - A L^{-\phi}$ with
$A$ and $\phi$ as fit parameters. The minimum of $\chi^2$ is
$1.7\times10^{-4}$ for $\bar c_{12}(\infty) = 0.54$.
For $\bar c_{12}(\infty) = 1$ we have $\chi^2 = 3.4\times10^{-4}$.
Our result $\bar c_{12}(\infty) <1$ should be tested further for
larger systems. The results $|\G_2(\infty)|>0$ and $\bar c_{12}(\infty) <1$
yield a non-trivial $P(q)$ \cite{HedPq}.

{\it Summary:} The ground states of the $\pm J$ short-range Ising 
spin glass have a hierarchical, tree-like structure. 
This structure is induced by correlated spin domains, which
are the cores of macroscopic zero energy excitations, taking
the system from one state-cluster to another.
This structure of  GS and the associated barriers has
some features of the MF picture, but is inconsistent with it,
since it lacks ultrametricity. 
It is, however, consistent with the FH scenario.

Note that evidence for low-energy macroscopic excitations has been found by
\cite{Krzakala00}; we presented here detailed statistics of these domains,
investigated their correlations and demonstrated that they produce
an hierarchical structure in state space. 

We thank the Germany-Israel Science Foundation for support, I. Kanter and
M. Mezard for helpful discussions, and N. Jan for hospitality to DS.



\begin{thebibliography}{99}
\bibitem{foot1} In fact we ensured to have 
equal numbers of positive and negative bonds.
\end{thebibliography}
\end{document}